\documentclass[twocolumn]{aastex7}
\usepackage{CJK}
\usepackage{multirow}
\usepackage{amsmath}

\newcommand{\bdv}[1]{\mbox{\boldmath$#1$}}

\newcommand{\mjup}{M_{\rm J}}
\newcommand{\msini}{m_{\rm p}\sin{i}}

\defcitealias{Zhu:2022}{Paper I}

\shorttitle{Giant Planet Intrinsic Multiplicities from HARPS \& CLS}
\shortauthors{Li \& Zhu}
\graphicspath{{./}{figures/}}
\begin{document}
\begin{CJK*}{UTF8}{gbsn}

\title{The Intrinsic Multiplicity Distribution of Exoplanets Revealed from the Radial Velocity Method. II. Constraints on Giant Planet Multiplicity from Different Surveys}

\correspondingauthor{Wei Zhu}
\email{weizhu@tsinghua.edu.cn}

\author[0009-0003-8853-4540]{Jiayin Li (李佳音)}
\affiliation{Department of Astronomy, Tsinghua University, Beijing 100084, China}
\email{jiayin-li25@mails.tsinghua.edu.cn}

\author[0000-0003-4027-4711]{Wei Zhu (祝伟)}
\affiliation{Department of Astronomy, Tsinghua University, Beijing 100084, China} 
\email{weizhu@tsinghua.edu.cn}

%% Note that the \and command from previous versions of AASTeX is now
%% depreciated in this version as it is no longer necessary. AASTeX 
%% automatically takes care of all commas and "and"s between authors names.

%% AASTeX 6.31 has the new \collaboration and \nocollaboration commands to
%% provide the collaboration status of a group of authors. These commands 
%% can be used either before or after the list of corresponding authors. The
%% argument for \collaboration is the collaboration identifier. Authors are
%% encouraged to surround collaboration identifiers with ()s. The 
%% \nocollaboration command takes no argument and exists to indicate that
%% the nearby authors are not part of surrounding collaborations.

%% Mark off the abstract in the ``abstract'' environment. 
\begin{abstract}
Compared to the commonly used planet occurrence rates, the multiplicity distribution of planets can be more useful in constraining the formation and evolution pathways of planetary systems. This work follows an earlier work of Zhu (2022) and derive the intrinsic multiplicity distribution of giant planets (with masses above Saturn mass) from two independent radial velocity (RV) surveys. In particular, we find that $(7.8\pm1.4\%, 2.3\pm1.2\%, 0.5^{+0.8}_{-0.3}\%)$ of Sun-like stars in the HARPS sample have $(1, 2, 3)$ giant planets within 10\,au, whereas $(7.3\pm2.8\%, 7.2\pm2.3\%, <1.3\%, 1.0^{+1.0}_{-0.6}\%)$ of Sun-like stars in the California Legacy Survey (CLS) have $(1, 2, 3, 4)$ giant planets within 10\,au. Here we have further cleaned the CLS sample and removed planet detections that were not discovered in the survey mode. The total fraction of Sun-like stars with at least one giant planet within 10\,au from the two samples are $10.6\pm1.2\%$ and $15.8\pm2.1\%$, respectively, and the difference may be accounted for by their different metallicity distributions. We briefly discuss the theoretical implications of our results. In particular, the inferred giant planet multiplicity distribution is inconsistent with most of the proposed theoretical models involving planet--planet scatterings, which predict too many multi-giant systems.
\end{abstract}

%% Keywords should appear after the \end{abstract} command. 
%% The AAS Journals now uses Unified Astronomy Thesaurus concepts:
%% https://astrothesaurus.org
%% You will be asked to selected these concepts during the submission process
%% but this old "keyword" functionality is maintained in case authors want
%% to include these concepts in their preprints.

%% From the front matter, we move on to the body of the paper.
%% Sections are demarcated by \section and \subsection, respectively.
%% Observe the use of the LaTeX \label
%% command after the \subsection to give a symbolic KEY to the
%% subsection for cross-referencing in a \ref command.
%% You can use LaTeX's \ref and \label commands to keep track of
%% cross-references to sections, equations, tables, and figures.
%% That way, if you change the order of any elements, LaTeX will
%% automatically renumber them.
%%
%% We recommend that authors also use the natbib \citep
%% and \citet commands to identify citations.  The citations are
%% tied to the reference list via symbolic KEYs. The KEY corresponds
%% to the KEY in the \bibitem in the reference list below. 

%%%%%%%%%%%%%%%%%%%%%%%%%%%%%%
\section{Introduction} \label{sec:intro}

Exoplanet multiplicity, namely the number of planets around the same host, provides important information about the formation and evolution of planetary systems \citep[e.g.,][]{Emsenhuber:2021, Bitsch:2023}. This quantity goes beyond the commonly used ``occurrence rates,'' as it measures not only the total number of planets for a given number of stars, but also how the planets are distributed among those stars. 

The intrinsic planet multiplicity that is directly relevant to the formation and evolution theories is, however, not easy to constrain from observations for multiple reasons. First, exoplanets around the same star may easily go beyond the detection limit of any single detection method \citep{Winn:2015, Zhu:2021}. Taking a solar system analog for an example, the planetary semi-major axes vary by a factor of $>100$ and the masses vary by a factor $\sim10^4$. Such a diverse planet family cannot be revealed by any of the currently known detection techniques. One promising approach is to combine the strength of multiple methods, and there have been some success in its practice (e.g., the inner--outer correlation studies, \citealt{Knutson:2014, Bryan:2016, ZhuWu:2018}), but it bears the limitation of individual involved methods and thus faces severe selection effects and/or small sample sizes. The multiplicity distribution is especially difficult to recover for the transit method, because the geometric requirement introduces a strong degeneracy between the multiplicity and inclination distributions \citep{Lissauer:2011, Tremaine:2012}, which cannot be easily broken without further information or additional assumptions \citep[e.g.,][]{Zhu:2018, Mulders:2018, He:2020}.

Another reason why the intrinsic multiplicity has rarely been studied is in the complexity of the statistical modeling. To derive the intrinsic multiplicity from the observations, one needs to correct for the detection efficiency of individual systems in the statistical sample, but this correction implicitly requires knowing the planet multiplicity (or the system architecture) as input. For comparison, to derive the planet occurrence rate, one only needs to estimate the detectability of individual planets, which can be done through the common injection--recovery exercise \citep[e.g.,][]{Cumming:2008, Mayor:2011}. To properly account for the detection biases into the derivation of the intrinsic planet multiplicity, a general statistical framework was established by \citet{Tremaine:2012}. The framework builds on the separability approximation,
\footnote{In fact, this assumption also applies to almost all statistical studies on exoplanet occurrence rates, even though it was never explicitly stated.}
namely that the distribution of planetary parameters is independent of the planet multiplicity and that the detections of planets around the same star are largely mutually independent. This separability approximation is valid is most but not all cases, as has been discussed in \citet{Tremaine:2012}.

An earlier work by \citet[hereafter Paper I]{Zhu:2022} applied the framework of \citet{Tremaine:2012} to the RV planet sample from the California Legacy Survey \citep[CLS,][]{Rosenthal:2021}. Focusing on the Sun-like star sample from CLS, \citetalias{Zhu:2022} reported the intrinsic multiplicity distribution of different types of planets, especially the giant planets (defined as those with minimum masses above $0.3\,\mjup$). One issue raised there is that the rate of hot Jupiters from CLS is substantially higher than the commonly accepted value ($\sim1\%$) by nearby a factor of $\sim 3$ \citep{Gan:2023}, raising the possibility that the CLS sample may suffer from additional selection biases that were not fully uncovered in its construction. As the planet occurrence rates from the CLS sample \citep{Fulton:2021} have been widely used for various purposes, it is necessary to revisit the issue, and that motivates the current work. 

In this work, we follow the method of \citetalias{Zhu:2022} and apply it to a different and independent RV sample \citep[namely HARPS,][]{Mayor:2011}. We also perform additional cleaning to the CLS sample to further reduce the impact of selection biases associated with that sample. We choose to focus only on the giant planet populations for multiple reasons. First, such massive planets typically produce statistically significant RV signals, so that their detections can indeed be treated as independent events, thus satisfying the separability approximation. Second, giant planets as the dominating objects in the planetary system are expected to be crucial in shaping the system architecture. These planets can affect the distribution of material in protoplanetary disks and dynamically influence other planets through migration and gravitational interactions \citep[e.g.,][]{Morbidelli:2016, Kley:2012, Bitsch:2023}. From this perspective, the intrinsic multiplicity of giant planets can provide insights not only into their formation and evolution, but also into the dynamics of other planets and even the history of the entire planetary system. Furthermore, it is widely accepted that giant exoplanets should have gone through a phase of dynamical 
instabilities, and different evolution models gave different predictions about their multiplicity distributions \citep[e.g.,][]{Juric:2008, Chatterjee:2008, Ford:2008, Wu:2011, Frelikh:2019}, to which the statistical results can be directly compared.

The paper is organized as follows. In Section \ref{sec:method}, we briefly review the statistical method used in this work. In Section \ref{sec:harps}, we present the inferred planet multiplicity distribution for the HARPS sample. In Section \ref{sec:cls}, we conduct additional selections to the CLS sample and obtain revised multiplicity results, which differ from those reported in \citetalias{Zhu:2022}. In Section \ref{sec:discussion}, we compare the results from the two samples and briefly discuss the implications to the dynamical evolution of giant planets.

%%%%%%%%%%%%%%%%%%%%%%%%%%%%%%
\section{Method Overview} \label{sec:method}

The method to derive the intrinsic multiplicity distribution follows closely that of \citetalias{Zhu:2022}. Here we provide a brief summary.

The observed multiplicity distribution $\bdv{N}$ is related to the intrinsic multiplicity vector $\bdv{F}$ via
\begin{equation} \label{eqn:nsf}
\bdv{N} = N_\star \bar{S} \bdv{F} .
\end{equation}
Here $N_\star$ is the number of surveyed stars, and $\bar{\mathcal{S}}$ is the average of the sensitivity matrix $\mathcal{S}$ over the whole stellar sample (with and without planet detections). For a star with a detection efficiency of $p$, the element of the sensitivity matrix, $S_{jk}$, which denotes the probability to yield $j$ planet detections out of $k$ planets, is given by \citep{Tremaine:2012}
\begin{equation} \label{eqn:s_matrix}
    S_{jk} (p) = \left\{
    \begin{array}{ll}
        \frac{k!}{j! (k-j)!} p^j (1-p)^{k-j} , &  j \leq k \\
        0 , & j > k
    \end{array} \right. .
\end{equation}
The average of $\mathcal{S}$ can be done in two different ways. The more accurate way is to take the average of the sensitivity matrices from individual stars, as is done in \citetalias{Zhu:2022} for the CLS Sun-like sample. This requires the knowledge of the sensitivity maps (or more precisely the detection efficiency parameters) of all stars in the statistical sample, which are usually not made public. A less accurate but more efficient way is to first use the averaged sensitivity map to derive the average detection efficiency parameter, $\bar{p}$, and then insert it into Equation~(\ref{eqn:s_matrix}) to derive $\bar{\mathcal{S}}$. This is going to be the approach we adopt for the HARPS sample.

A key parameter in the derivation of the intrinsic multiplicity distribution is the parameter $p$. By definition, it is the averaged detection probability of the star to a typical planet within the predefined parameter space weighted by the intrinsic planet distribution \citepalias{Zhu:2022}. Therefore, one needs to first estimate the intrinsic planet distribution (namely occurrence rate of planets as a function of planetary properties) in order to know the value of $p$. The implicit assumption in this step is that the multiple planet detections around the same star are (largely) unrelated. As we focus on RV giant planets, which usually produce significant signals even in the presence of additional giant planets, this assumption generally holds.

The intrinsic multiplicity distribution is given by
\begin{equation}
\bdv{F} = \frac{1}{N_\star} S^{-1}(p) \bdv{N}.
\end{equation}
In reality, the values and uncertainties are obtained via the maximum likelihood method. We keep the maximum intrinsic multiplicity at the maximum observed multiplicity. The inclusion of even higher multiplicities would modify only slightly the derived intrinsic multiplicity distribution, and it has even less impact on the derived quantities like $F_{\rm p}$ and $\bar{m}_{\rm p}$, as has been shown in \citetalias{Zhu:2022}.

With the intrinsic multiplicity vector, the frequency of planetary systems is known through
\begin{equation}
    F_p = \sum_{k=1} F_k ,
\end{equation}
and the average multiplicity is then
\begin{equation}
    \bar{m}_p = \frac{\bar{n}_p}{F_p} .
\end{equation}

%%%%%%%%%%%%%%%%%%%%%%%%%%%%%%
\section{Giant Planet Multiplicity from HARPS} \label{sec:harps}

\begin{figure}
    \centering
    \includegraphics[width=0.95\columnwidth]{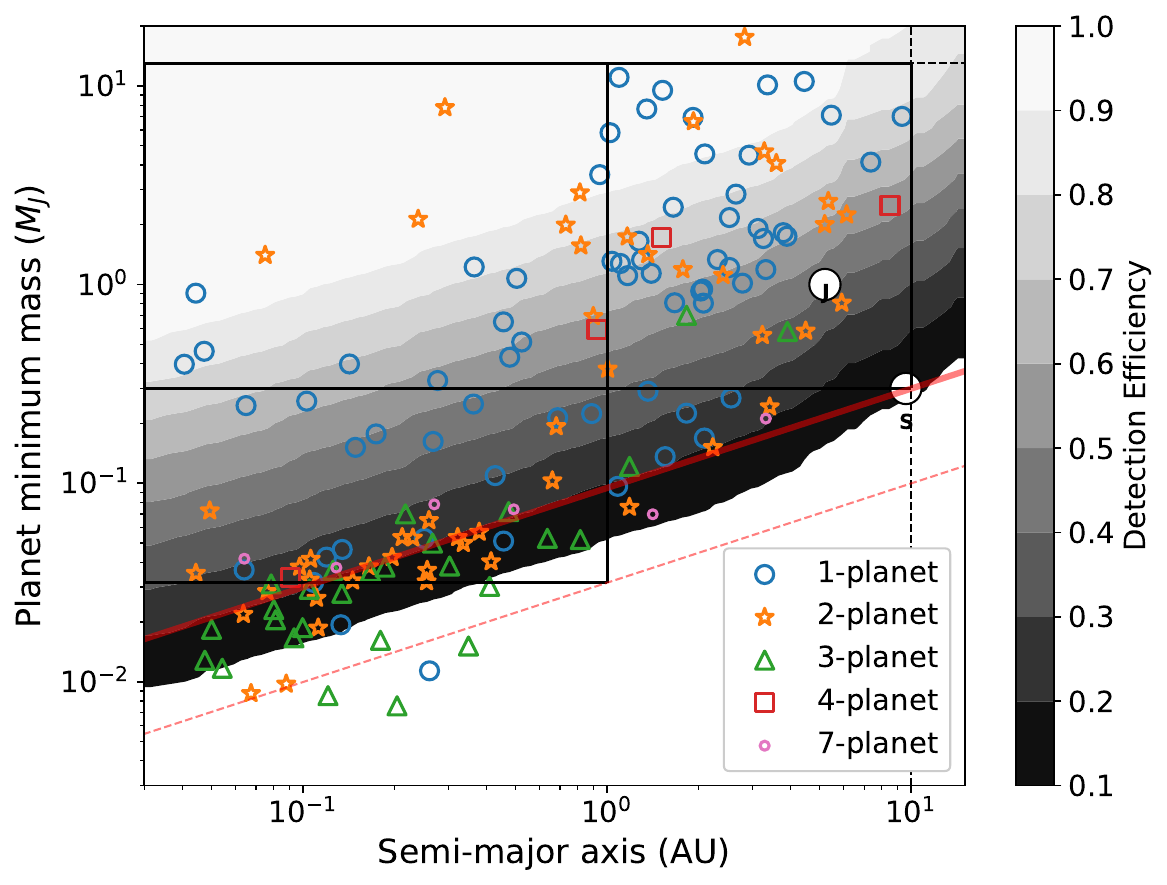}
    \caption{The distribution of 822 HARPS sample stars in the minimum mass vs. semi--major axis plane. Planets with different observed multiplicities are differentiated with different labels and colors. The contours represent the survey completeness, with lighter shades indicating higher sensitivity and darker shades indicating lower sensitivity. Both based on the data from \citet{Mayor:2011}. The vertical dashed line indicates the outer boundary (10~au) that is considered in this work, and the horizontal dashed line indicates the upper mass limit (13~M$_{\mathrm{J}}$). The positions of Jupiter and Saturn are also shown. The regions of different planet categories defined in Section~3 are outlined by the black boxes. The red solid and dashed lines denote the RV semi-amplitudes of 3 and 1~m\,s$^{-1}$, respectively.} 
    \label{fig:harps-sample}
\end{figure}

\begin{figure}
    \centering
    \includegraphics[width=0.95\columnwidth]{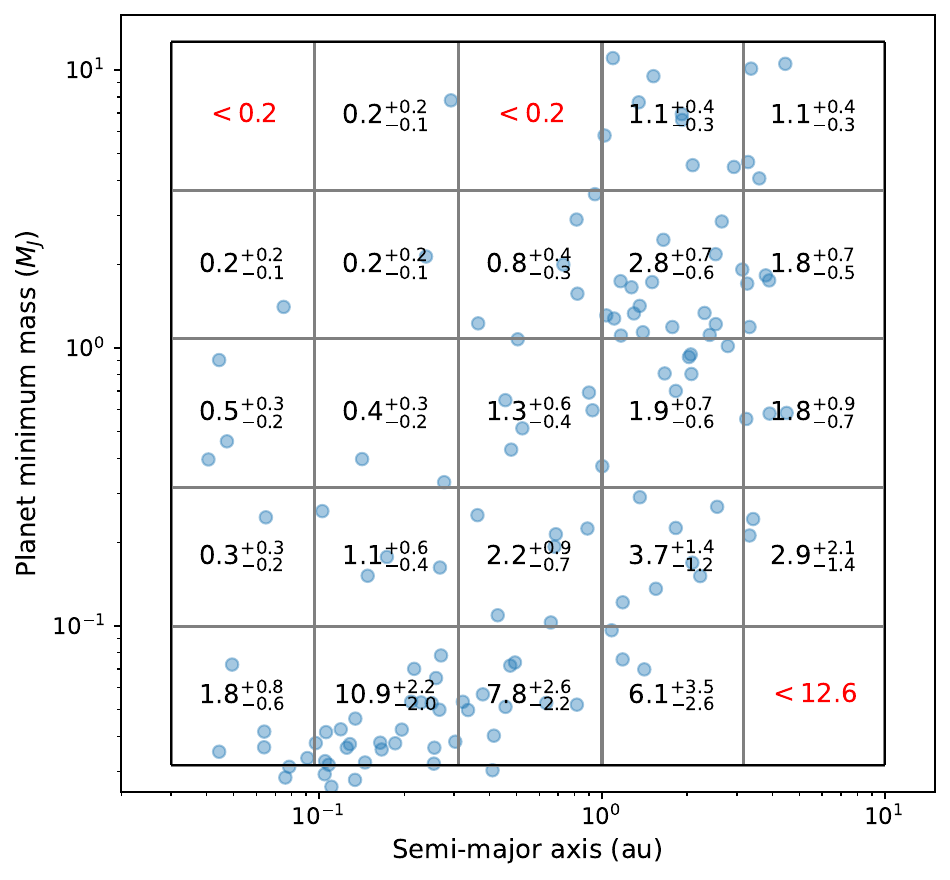}
    \caption{This figure shows in blue dots the planets detected in the HARPS sample whose properties fall within our chosen parameter space. The values within each grid cell give the number of such planets per 100 stars. Error bars correspond to the 68\% confidence interval, whereas for grid cells with fewer than two planet only the 95\% upper limit is shown. The integrated planet frequencies are $\bar{n}_{\mathrm{p}} = 0.036, \approx 0.104, \approx 0.140$ for our definitions of close-in, cold, and all giant planets, respectively.}
    \label{fig:harps-rate}
\end{figure}

\begin{deluxetable}{cccccccc}
\tablecaption{The Observed Multiplicity Distribution and the Estimated Mean Detection Probability $p$ for Different Types of Planets in the HARPS sample.
\label{tab:harps_multiplicity}}
\tablehead{
\colhead{} & \colhead{Close-in giant} & \colhead{Cold giant} & \colhead{All giant}
}
\startdata
    $N_0$ & 803 & 776 & 763\\
    $N_1$ & 19 & 41 & 49\\
    $N_2$ & 0 & 5 & 9\\
    $N_3$ & 0 & 0 & 1\\
    $\bar{p}$ & 0.64 & 0.60 & 0.61\\
\enddata
\tablecomments{We define three types of giant planets:
Close-in giants ($a<1\,$au, 0.3--13\,$\mjup$),
Cold giants (1--10\,au, 0.3--13\,$\mjup$),
and All giants ($a<10\,$au, 0.3--13\,$\mjup$).
}
\end{deluxetable}

\begin{deluxetable}{cccc}
\tablecaption{Intrinsic Multiplicities for Different Classes of Giant Planets in the HARPS sample. The definitions of the three types of planets are the same as in Table~\ref{tab:harps_multiplicity}. \label{tab:harps_result}}
\tabletypesize{\small}
\tablehead{
\colhead{} & \colhead{Close-in giant} & \colhead{Cold giant} & \colhead{All giant}
}
\startdata
    $F_1$ & $3.7\pm0.6\%$ & $6.8\pm1.2\%$ & $7.8\pm1.4\%$ \\
    $F_2$ & $<0.1\%$ & $1.7\pm0.7\%$ & $2.3\pm1.2\%$\\
    $F_3$ & \nodata & \nodata & $0.5^{+0.8}_{-0.3}\%$\\
    \hline
    $F_{\rm p}$ & $3.6\pm0.6\%$ & $8.7\pm1.1\%$ & $10.6\pm1.2\%$\\
    $\bar{n}_{\rm p}$ & $0.04\pm0.01$ & $0.10\pm0.01$ & $0.14\pm0.02$\\
    $\bar{m}_{\rm p}$ & $1.0$ & $1.20\pm0.11$ & $1.32\pm0.13$\\
\enddata
\end{deluxetable}

\begin{deluxetable}{cccccccc}
\tablecaption{The Estimated Mean Detection Probability for Different Classes of Giant Planets
\label{tab:different_pbar}}
\tablehead{
\colhead{} & \colhead{Close-in giant} & \colhead{Cold giant} & \colhead{All giant}
}
\startdata
    $\bar{p}$ & 0.64 & 0.60 & 0.61\\
    $\bar{p}_2$ & 0.57 & 0.47 & 0.49\\
    $\bar{p}_{\rm inf}$ & 0.76 & 0.66 & 0.68\\
\enddata
\tablecomments{Here, the first and second rows both adopt Equation~(\ref{eq:probability}) for the calculation. The first row follows the grid division shown in Figure~\ref{fig:harps-rate}, while the second row employs a refined sub-grid based on Figure~\ref{fig:harps-rate} to compute $\bar{p}$. The third row provides $\bar{p}$ calculated using the IDEM method.
}
\end{deluxetable}

The HARPS sample contains 155 RV planet detections/candidates out of 822 Sun-like stars \citep{Mayor:2011}. The status of a few claimed planet detections in the \citet{Mayor:2011} sample has not been confirmed by the standard of the NASA Exoplanet Archive \citep[NEA,][]{Christiansen:2025}.
\footnote{See \url{https://exoplanetarchive.ipac.caltech.edu/docs/removed_targets.html} for more details.}
However, this concern applies primarily to the lower-mass planets. The only system with giant planet(s) in the ambiguous status is HD~114386. \citet{Mayor:2011} claimed two giant planet detections with $\msini\approx119\,M_\oplus$ and $377\,M_\oplus$ at $P\approx445\,$d and $1046\,$d, respectively, whereas NEA only adopts the original planet detection with $\msini \approx 273\,M_\oplus$ and $P \approx 938\,$d from \citet{Mayor:2004}. Given that there are in total 70 giant planet detections out of 59 planetary systems in the \citet{Mayor:2011} sample, the impact is negligible regarding whether correcting or not this ambiguous system. We choose to adopt the parameters of \citet{Mayor:2011} to be consistent with the previous analysis \citep[see also][]{Fernandes:2019}.

Information about the HARPS planet detections and the averaged detection efficiency curves are obtained from the \texttt{epos} package \citep{Mulders:2018} and illustrated in Figure~\ref{fig:harps-sample}. The observed multiplicity distributions of the three types of giant planets are given in Table~\ref{tab:harps_multiplicity}.

To derive the average detection efficiency parameter $p$ that is needed for the multiplicity analysis, we first derive the planet occurrence rates in the $\msini$ vs.\ $a$ plane. The derived rates for the set of grid cells are shown in Figure~\ref{fig:harps-rate}. In doing so we have assumed that the planetary distribution within the individual grid is uniform and that the detection is a binomial process (see \citealt{Zhu:2024} for further details).

The occurrence rate of planets defined in a certain parameter space can be obtained if one sums up the numbers of enclosed grid cells. We use $\bar{n}_p^{\rm dir}$ to denote this directly derived planet rate. For our definitions of close-in giants, cold giants, and all giants, these values are $\approx0.036$, $\approx0.104$, and $\approx0.140$, respectively. We can then further estimate the values of the average detection efficiency parameter for these different types of planets through \citepalias{Zhu:2022}
\begin{equation}
    \boldsymbol{\bar{p} = \frac{N_p/N_\star}{\bar{n}_p}} .
    \label{eq:probability}
\end{equation}
These values are listed as the last line in Table~\ref{tab:harps_multiplicity}. We then derive the intrinsic multiplicity distribution of these three types of planets via Equation~(\ref{eqn:nsf}), and the results are given in Table~\ref{tab:harps_result}. Below we discuss them in some details:

\begin{itemize}
    \item \emph{Close-in giants ($a<1\,$au, 0.3--13\,$\mjup$)}. The HARPS sample contains 19 such planets, none of which have companions of the same type. With an average detection efficiency of $\bar{p}=0.64$, it reveals that $3.7 \pm 0.6\%$ of Sun-like stars host one close-in giant planet, and that the rate of systems with two such planets is $<0.1\%$. The average multiplicity of such planets is unity. In particular, the occurrence rate of hot ($<0.1\,$au) giants is $\sim$0.7\%, in general agreement with previous estimates based on RV samples \citep[e.g.,][]{Wright:2012}.
    \item \emph{Cold giants (1--10\,au, 0.3--13\,$\mjup$)}. The HARPS sample contains 51 such planet, 10 of which are in systems with two such planets. With $\bar{p}=0.60$, the results shows that $6.8\pm1.2\%$ and $1.7\pm0.7\%$ of Sun-like stars in this sample have one and two cold giants, respectively. The total fraction of Sun-like stars with such planets is then $8.7\pm1.1\%$, and the average multiplicity is $\approx 1.2$.
    \item \emph{All giants ($a<10\,$au, 0.3--13\,$\mjup$)}. The HARPS sample contains 70 such planets, with (49, 9, 1) having $j=1$, 2, and 3 planet detections, respectively. Using $\bar{p}=0.61$, we find that there are $7.8\pm1.4\%$, $2.3\pm1.2\%$, and $0.5^{+0.8}_{-0.3}\%$ of Sun-like stars have one, two, and three giant planets within 10\,au. These sum up to $10.6\pm1.2\%$ of Sun-like stars with at least one giant planet within 10\,au, and that every giant planet system has on average $1.32\pm0.13$ such planets.
\end{itemize}

When applied to the same parameter space, our method yields consistent results with previous works. For example, applying the method to the giant planets within 5\,au (or equivalently orbital period $P\lesssim10\,$yr), we find the frequency of planetary systems to be $10.1\pm1.5\%$. This value is consistent with the early result of \citet{Mayor:2011}, which used the detectability of a single planet with the largest signal to derive the fraction of stars with planets. The two methods yield similar results as long as the average multiplicity is close to unity, which is the case for giant planets within 5\,au ($\bar{m}=1.18\pm0.09$), but they may deviate substantially if the average multiplicity is high. Compared to the method of \citet{Mayor:2011}, our method is able to properly correct for this multiplicity effect and furthermore directly give the intrinsic multiplicity distribution.

The above derivation has been based on the average detection efficiency parameter derived from Figure~\ref{fig:harps-rate}. While the grid in that figure has been chosen to capture the main feature of the underlying giant planet distribution (e.g., the rapid increase in occurrence rate at ~1\,au, \citealt{Cumming:2008, Petigura2018, Wittenmyer2020, Lagrange2023}), the number of grid points affects the exact value of $\bar{p}$ and thus the derived intrinsic multiplicity distribution. Table~\ref{tab:different_pbar} compares the values of $\bar{p}$ derived with three different methods for the three types of giant planets. Specifically, $\bar{p}_2$ is estimated by further refining the grids in Figure~\ref{fig:harps-sample} by factors of two along both dimensions, and $\bar{p}_{\rm inf}$ is the estimated by evaluating and inverting the sensitivity only at positions of the observed planets. The latter is effectively the inverse detection efficiency method (IDEM, \citealt{Foreman-Mackey:2014}) that is widely used for occurrence calculations \citep[e.g.,][]{Cumming:2008, Mayor:2011, Wittenmyer2020}. The different estimations give a systematic uncertainty at the level of $<20\%$ in $\bar{p}$, which is smaller than the statistical uncertainties on the derived intrinsic multiplicities.

\begin{figure}
    \centering
    \includegraphics[width=0.95\columnwidth]{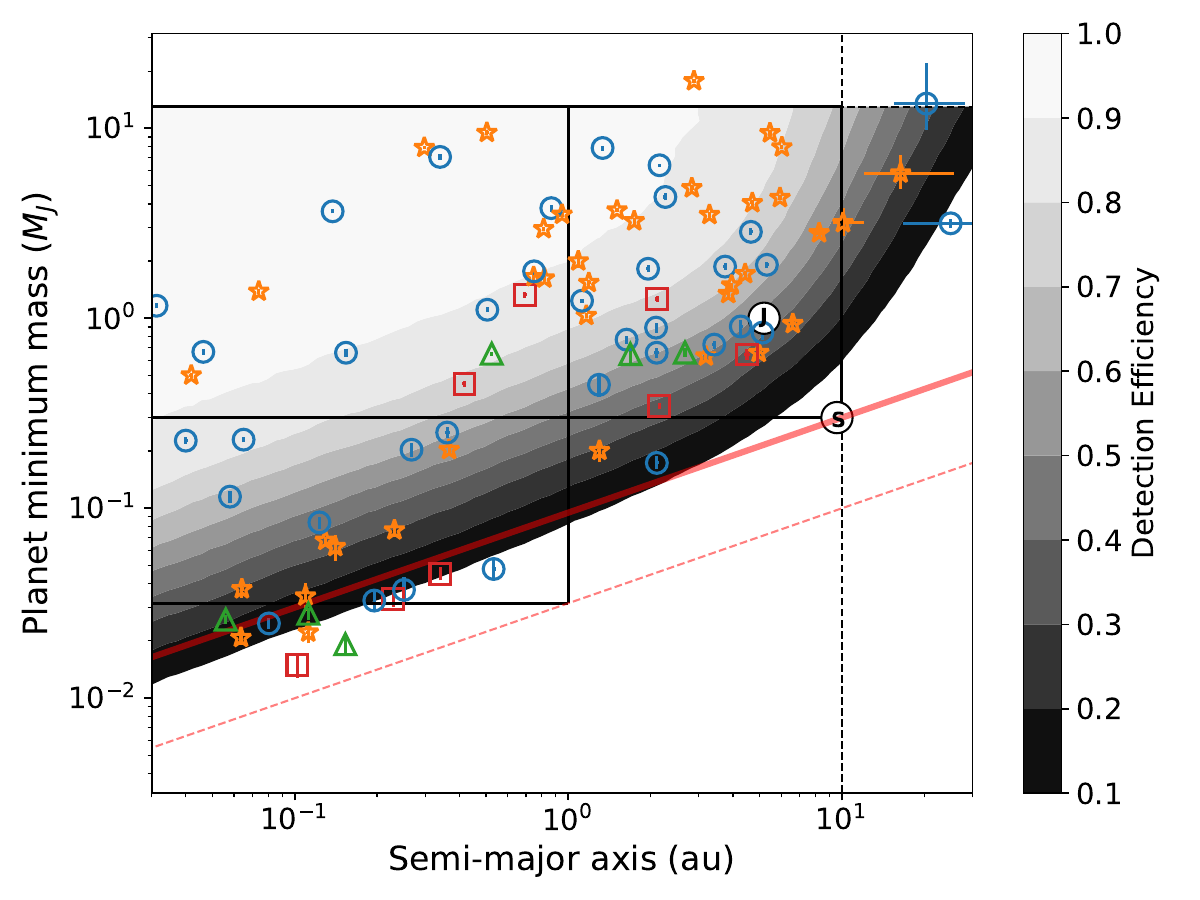}
    \caption{The distribution of the remaining 351 CLS sample stars in the minimum mass vs. semimajor axis plane after selection. The meanings of the markers, background contours, and various lines are the same as described in the caption of Figure~\ref{fig:harps-sample}.}
    \label{fig:cls-sample}
\end{figure}

\begin{figure}
    \centering
    \includegraphics[width=0.95\columnwidth]{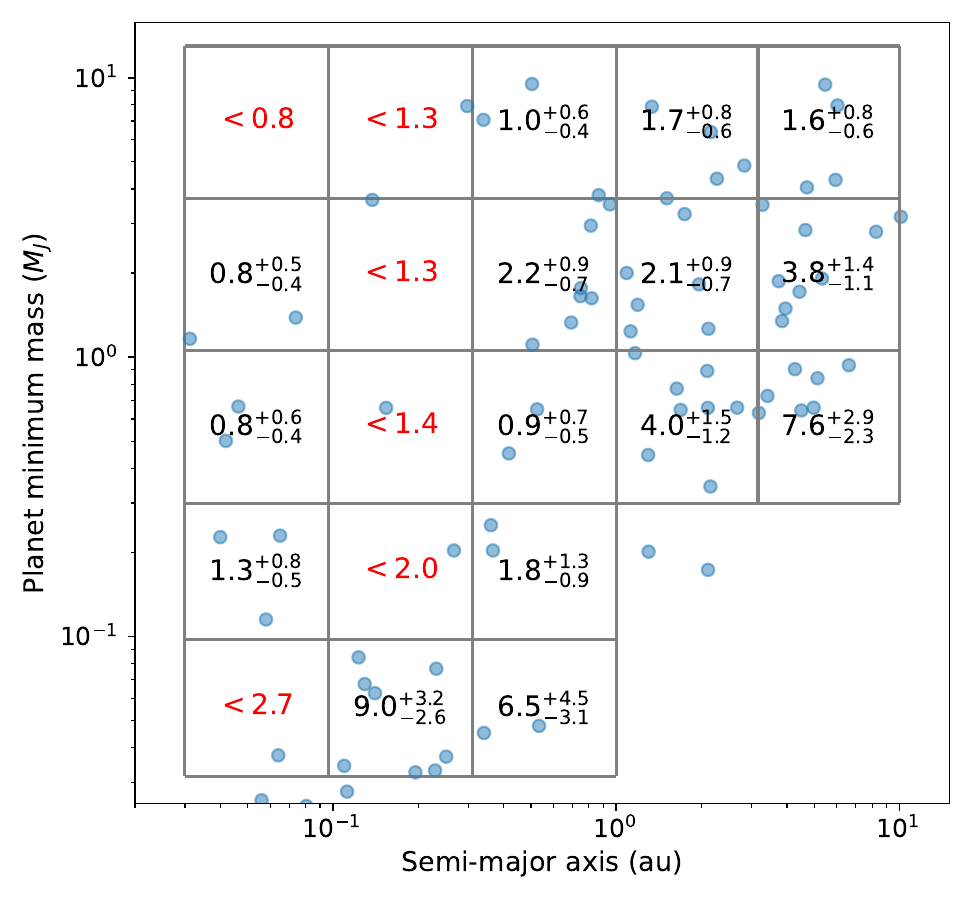}
    \caption{This figure shows in blue dots the planets detected in the less biased CLS sample whose properties fall within our chosen parameter space. The values shown in the grid and their meanings are the same as described in the caption of Figure~\ref{fig:harps-rate}.}
    \label{fig:cls-rate}
\end{figure}

\begin{deluxetable}{cccccccc}
\tablecaption{The Observed Multiplicity Distribution and the Estimated Mean Detection Probability $p$ for Different Types of Planets in the less biased CLS sample. The definitions of the three types of planets are the same as in Table~\ref{tab:harps_multiplicity}.
\label{tab:cls_multiplicity}}
\tablehead{
\colhead{} & \colhead{Close-in giant} & \colhead{Cold giant} & \colhead{All giant}
}
\startdata
    $N_0$ & 333 & 317 & 309\\
    $N_1$ & 17 & 29 & 29\\
    $N_2$ & 1 & 5 & 11\\
    $N_3$ & 0 & 0 & 1\\
    $N_4$ & 0 & 0 & 1\\
    $\bar{p}$ & 0.90 & 0.54 & 0.63\\
\enddata
\end{deluxetable}

\begin{deluxetable}{cccccccc}
\tablecaption{Intrinsic Multiplicities for Different Classes of Giant Planets in the less biased CLS sample. The definitions of the three types of planets are the same as in Table~\ref{tab:harps_multiplicity}.}\label{tab:cls_result}
\tabletypesize{\small}
\tablehead{
\colhead{} & \colhead{\multirow{2}{*}{Close-in giant}} & \colhead{\multirow{2}{*}{Cold giant}} & \colhead{\multirow{2}{*}{All giant}} \\
\colhead{} & \colhead{} & \colhead{} & \colhead{}
}
\startdata
    $F_1$ & $5.4\pm0.9\%$ & $10.9\pm3.1\%$ & $7.3\pm2.8\%$ \\
    $F_2$ & $0.3^{+0.5}_{-0.2}\%$ & $4.6\pm2.1\%$ & $7.2\pm2.3\%$\\
    $F_3$ & \nodata & \nodata & $<1.3\%$\\
    $F_4$ & \nodata & \nodata & $1.0^{+1.0}_{-0.6}\%$\\
    \hline
    $F_{\rm p}$ & $5.7\pm1.1\%$ & $15.7\pm2.4\%$ & $15.8\pm2.1\%$\\
    $\bar{n}_{\rm p}$ & $0.06\pm0.01$ & $0.21\pm0.03$ & $0.26\pm0.04$\\
    $\bar{m}_{\rm p}$ & $1.06\pm0.07$ & $1.31\pm0.16$ & $1.66\pm0.21$\\
\enddata
\end{deluxetable}

\section{A less Biased CLS Sample} \label{sec:cls}

An issue in the CLS sample, first identified by \citetalias{Zhu:2022}, is that the resulting hot Jupiter occurrence around Sun-like stars is nearly a factor of three higher than the widely accepted value \citep[e.g.,][]{Wright:2012, Gan:2023}, which signals some unknown biases in the sample construction of CLS. As close-in giant planets preferentially have cold giant companions \citep[e.g.,][]{Knutson:2014, Bryan:2016, Zink2023}, the high occurrence rate of hot Jupiters may well affect the occurrence and multiplicity of cold giants as well as all giants.

To come up with a less biased CLS Sun-like star sample, we have implemented the following cuts to the original CLS sample: 
\begin{enumerate}
    \item Only targets from the original Keck Planet Search are considered. These are the stars that were added into the sample between 1996 and 2004 \citep{Cumming:2008}. This cut allows us to compare results with that of \citet{Cumming:2008}, which has long been considered one of the standard works in the giant planet occurrence study. After this selection, 605 stars remain in the sample.
    \item Sun-like stars are selected based on the stellar $B-V$ color and $V$ band absolute magnitude \citep{Wright:2012}, similar to the selection criterion used in HARPS sample \citep{Mayor:2011}. Specifically, we require $B-V<1.2$ and $M_V$ to within $2.5$\,mag of the empirical main sequence of \citet{Wright:2005}, which together remove M dwarfs and evolved stars from the sample. After this step, 455 stars remain.
    \item A magnitude cut ($G<8$) is imposed to remove faint stars. As has been argued in \citet{Wright:2012}, this cut is necessary in order to remove targets that were added because there had been additional knowledge suggesting the existence or probable existence of planet(s). One example is BD-10~3166, which has $G=9.8$ and is the faintest hot Jupiter host in the CLS sample. This star was observed because of its very high bulk metallicity ([Fe/H]$\approx0.5$, \citealt{Butler:2000}).  After applying this magnitude cut, 358 stars remain.%  {\wei (how many remain after this step?)}
    \item For giant planet detections whose host stars pass the above cuts, we manually checked their original discovery paper against the first observation date by CLS, to identify and remove targets that had been known prior to the CLS observations. The following six targets were removed by this criterion: HD~12661, HD~130322, HD~143761, HD~179949, HD~74156, and HD~28185.
\footnote{Observations from Keck on HD~12661 were started after the Lick observations suggested the presence of a planetary companion \citep{Fischer:2001}. For HD~130322, HD~143761, and HD~28185, Keck observations started after their discoveries had been reported by other RV teams \citep{Udry:2000, Noyes:1997, Santos:2001}. Similarly, HD~179949 was first found to host planet by AAT, and the Keck observations were ``confirmatory'' \citep{Tinney:2001}. For HD 74156, although the discovery paper was not published until 2004, the planet discovery was already announced in an ESO press release on April 4, 2001, which is before the first Keck observation (April 8, 2001).}
    Additionally, we also checked the current status of all giant planet detections in NEA and delete HD~33636 on the basis that the reported detection is no longer a planet \citep{Bean:2007}. In total, we removed seven targets in this step.
\end{enumerate}

Figure~\ref{fig:cls-sample} illustrates the CLS sample after the above selections. Out of the 605 Sun-like stars initially monitored by the Keck survey, 351 remain after applying our filtering criteria. Compared with the 822 targets in the HARPS program, this refined sample includes roughly half as many stars. The corresponding observed multiplicity distribution is summarized in Table~\ref{tab:cls_multiplicity}.

We performed a similar statistical analysis to the cleaned CLS sample. A notable difference from the analysis done in Section~\ref{sec:harps} for the HARPS sample is that \citet{Rosenthal:2021} provided results from their injection--recovery simulations for all 719 CLS targets, including the 351 stars retained in our refined sample. These simulations enable a direct evaluation of the detection sensitivity for individual stars in the sample, which provides a more reliable evaluation of the sensitivity matrix \citepalias{Zhu:2022}.

As shown in Figure~\ref{fig:cls-rate}, giant planets in the refined CLS sample appear more common than in the HARPS results. For close-in giants, cold giants, and all giants, the directly inferred mean occurrence rates $\bar{n}_p^{\rm dir}$ are approximately 0.054, 0.204, and 0.258, respectively.

The resulting average detection efficiencies for the three planet categories are listed in the last row of Table~\ref{tab:cls_multiplicity}. We then apply Equation~(\ref{eqn:nsf}) to measure the intrinsic multiplicity distribution of giant planets in the refined CLS sample, and the results are summarized in Table~\ref{tab:cls_result}.
The meanings of these numbers are the same as those in Table~\ref{tab:harps_result}, which are explained in Section~\ref{sec:cls}, so they are not repeated here. The rates of giant planets derived from the refined CLS sample are systematically higher than those from HARPS, and we discuss in Section~\ref{sec:comparison} the plausible explanation.

We now compare the occurrences and multiplicities of giant planets derived from the CLS sample before and after the clean-ups. We find that the fractions of Sun-like stars having at least one giant planets in ranges of $<1\,$au, 1--10\,au, and $<10\,$au in the refined CLS sample are $5.7\pm1.1\%$, $15.7\pm2.4\%$, and $15.8\pm2.1\%$, respectively. For comparisons, the original CLS Sun-like sample gave $7.2\pm1.6\%$, $17\pm3\%$, and $19.2\pm2.8\%$ for the same planet definitions \citepalias{Zhu:2022}. These systematic differences do confirm that the general CLS sample is biased toward stars with planet detections.

Another useful comparison is with \citet{Cumming:2008}. By construction, our refined CLS sample is very similar to the sample that was used in \citet{Cumming:2008}, so we expect to have very similar rates. This is indeed the case. \citet{Cumming:2008} reported the fraction of Sun-like stars with giant planets within 10\,au to be $14\pm2\%$, in general agreement with our result. Similar to \citet{Mayor:2011} for the HARPS sample, \citet{Cumming:2008} only included one planet with the highest Doppler amplitude in the multi-planet systems, which prevented them from deriving the multiplicity distribution of planets or the average multiplicity.

%%%%%%%%%%%%%%%%%%%%%%%%%%%%%%
\section{Discussion} \label{sec:discussion}
\subsection{HARPS vs.\ CLS and the Role of Metallicity} \label{sec:comparison}

\begin{figure*}
    \centering
    \includegraphics[width=0.95\textwidth]{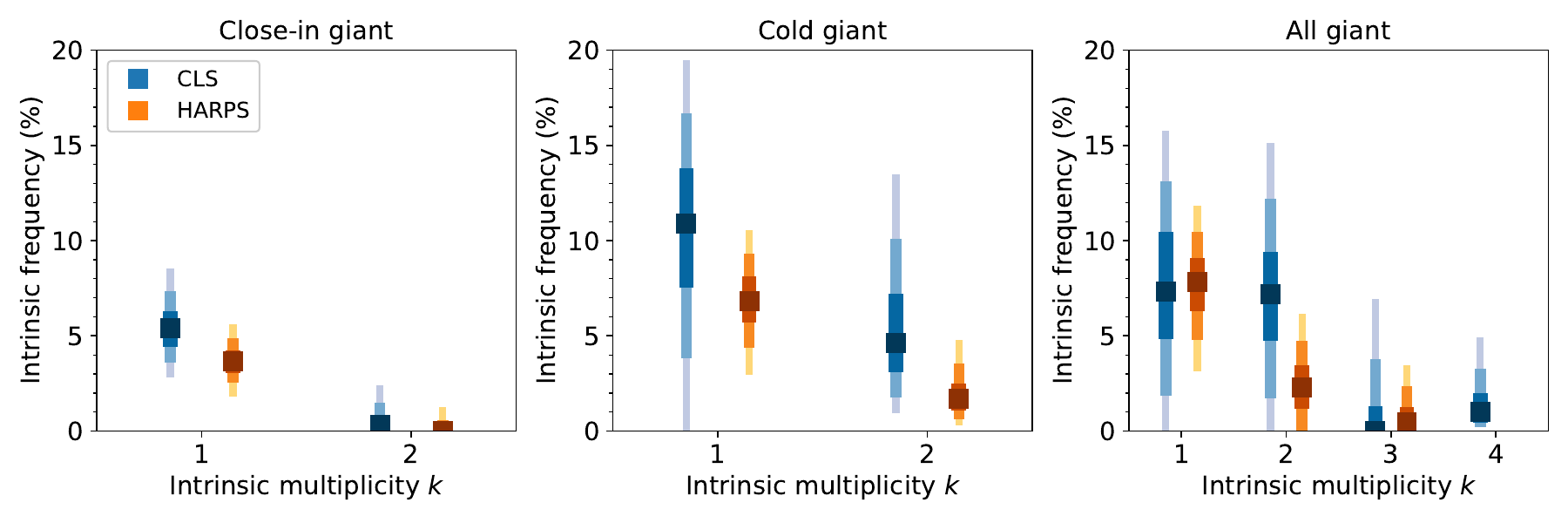}
    
    \caption{Illustrations of the derived intrinsic planet multiplicity distributions for the three chosen planet classes from the cleaned CLS (blue) and HARPS (orange) datasets. 
    The maximum-likelihood solutions and 1$\sigma$--3$\sigma$ confidence intervals are indicated by the colored bars with increasing transparency.}
    \label{fig:intrinsic_frequencies}
\end{figure*}

\begin{figure}
    \centering
    \includegraphics[width=0.95\columnwidth]{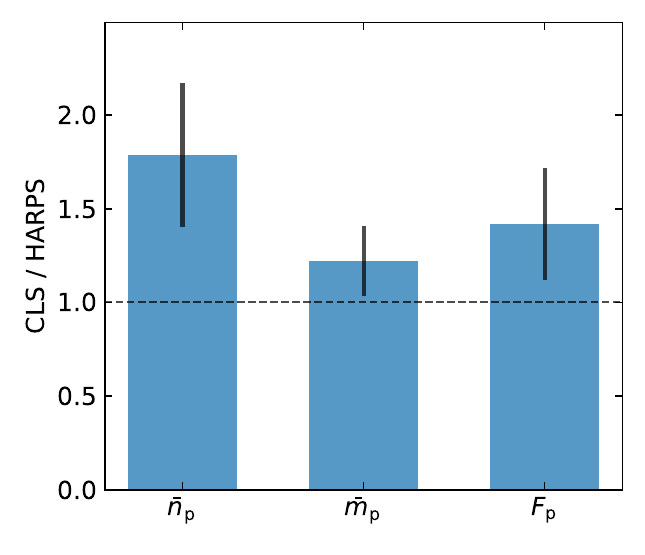}
    \caption{Ratios of CLS to HARPS values for average number of planets per star ($\bar n_{\rm p}$), average multiplicity ($\bar m_{\rm p}$), and the fraction of stars with planets ($F_{\rm p}$). The values are taken for the ``All giant'' population. Error bars indicate 1$\sigma$ uncertainties. The dashed line at unity marks equal values between the two samples.}
    \label{fig:cls_vs_harps}
\end{figure}

\begin{figure}
    \centering
    \includegraphics[width=0.95\columnwidth]{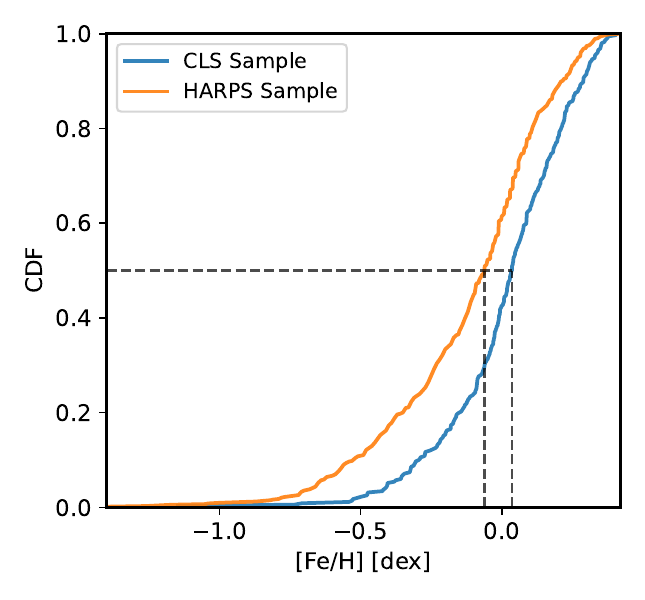}
    \caption{Comparison of the stellar metallicity distributions of the HARPS and refined CLS samples. The vertical dashed lines mark the mean values of each sample. The refined CLS sample is systematically more metal-rich, with a mean offset of approximately $0.1$ dex.}
    \label{fig:metallicity}
\end{figure}

In the previous sections, we have presented the multiplicity analysis of giant planets for both the HARPS sample and the refined CLS sample. Direct comparisons between the two samples are illustrated in Figures~\ref{fig:intrinsic_frequencies} and \ref{fig:cls_vs_harps}. The derived rates are consistent with the previous works for either sample, but the two samples remain statistically inconsistent, despite our additional selections imposed on the original CLS sample to mitigate the sample biases as much as possible.  As the detailed information about the constructions of individual samples is unavailable, our ability to perform a rigorous comparison of the selection functions is limited.

Instead, We seek to understand the origin of the discrepancy between the two samples, under the assumption that it is not due to any further selection biases of either sample. Since the occurrence rate of giant planet systems, especially the close-in ones, is known to correlate with the bulk metallicity of host stars \citep[e.g.,][]{Santos2001, Santos:2004, Fischer:2005}, it makes sense to compare the mean stellar metallicities of HARPS and CLS samples. As shown in Figure~\ref{fig:metallicity}, the cleaned CLS sample exhibits a systematically higher metallicity by approximately $0.1$\,dex relative to the HARPS sample. Using the empirical giant planet--metallicity relation of \citet{Fischer:2005}, we expect a difference in giant planet frequencies to be $10^{2\Delta[\mathrm{Fe}/\mathrm{H}]} \approx 1.6$, 
which is in good agreement with the difference in the giant planet frequencies inferred from our multiplicity measurements (Figure~\ref{fig:cls_vs_harps}). 

If the difference between the two samples is indeed caused by the difference in the stellar metallicity, then our results have additional implications. Early studies on the giant planet--metallicity correlation have mainly focused on relatively close-in planets, because of the limited RV baseline back then. According to our results, the frequency of close-in ($<1\,$au) giant is increased by a factor of $\sim1.6$ from HARPS to CLS, and the frequency of cold (1--10\,au) giant is increased by a factor of $\sim1.8$. The two factors are statistically similar, suggesting that the metallicity enhancement not only applies to close-in giants, but to all giants within (at least) 10\,au. It also suggests that the close-in giants do not require an even more metal-rich environment than do cold giants, thus putting constraints on the in situ formation models of giants at close-in separations (see \citealt{Dawson:2018} and references therein), although one should also be cautious about the evolutionary effects like migration and dynamical scattering after the initial formation.

The occurrence rates of giant planets ($\bar{n}_{\rm p}$) seem to increase more from HARPS to CLS than do the occurrence rates of giant planet systems ($F_{\rm p}$): $\bar{n}_{\rm p}$ increases by factor of $\sim2.1$ for cold giants and $\sim1.9$ for all giants within 10\,au. This is presumably due to the fact that the average multiplicities of giant planets also seem to increase from HARPS to CLS: $\bar{m}_{\rm p}$ changes from $1.20\pm0.11$ to $1.31\pm0.16$ for cold giants and from $1.32\pm0.13$ to $1.66\pm0.21$ for all giants within 10\,au. The evidence remains marginal in the statistical sense, but if confirmed, it would suggest that the higher-metallicity systems do form and retain, after the dynamical evolution, more giant planets \citep[see also][]{Rosenthal:2024}. This behavior of the giant planet is different from the behavior of the smaller planets, namely super Earths, for which the average multiplicity seems to flatten or decrease around metal-rich stars \citep{Zhu:2019}. This feature deserves further investigations with larger and more uniform datasets.

The derived intrinsic multiplicity distributions of three chosen giant planet populations are illustrated in Figure~\ref{fig:intrinsic_frequencies}. Except for the overall lower rates of the HARPS sample, which may potentially be related to its more metal-poor stellar sample, there are a few common features to notice. First, giant planets do not usually have companions of the same type, especially the close-in ($<1\,$au) ones. Specifically, $\sim5\%$ of Sun-like stars have one giant planet within 1\,au, but $\lesssim 0.5\%$ of these stars have two such planets. Therefore, systems such as TIC~118798035, which has two transiting close-in giants \citep{Brahm:2025}, are intrinsically rare. Another common feature is that high ($k\ge3$) multiples are very rare too, even for the ``All giant'' population. Systems with at least three giant planets within $10\,$au contribute no more than $\sim10\%$ of all giant planet-hosting stars.

\subsection{Average Multiplicity of Giant Planet Systems}

As shown in Tables~\ref{tab:harps_result} and \ref{tab:cls_result}, the intrinsic multiplicity of giant planets in either the HARPS ($\sim 1.3$) or the refined CLS ($\sim 1.7$) sample is below two, indicating that giant planets rarely have companions of similar type in their systems.

The above result is limited to giant planets within 10\,au, because of the limited RV time baseline. To assess the impact of this somewhat subjective definition, which could potentially have neglected outer giants beyond 10\,au, we examined the number of host stars showing an obvious linear trend in their RV data. According to \citet{Rosenthal:2021}, among all CLS systems that have giant planets, only six exhibited a linear trend in their RV data. If we conservatively assume that all these linear trends are due to undetected giant planets (rather than binary companions or other effects), the average multiplicity would increase by no more than $1/7$, given that there are 42 giant-planet host stars in the refined CLS sample. This yields an enhanced giant planet multiplicity $\bar{m}_p \lesssim 1.9$ out to $\sim30\,$au, which still stays below two. The theoretical implication of it will be discussed in the next subsection. 
\subsection{Theoretical Implications}

Extra-solar giant planets typically have large ($e>0.1$) orbital eccentricities that cannot be explained by planet--disk interactions under normal conditions \citep{Duffell:2015}, and the leading explanation is that these planets have undergone planet--planet scatterings after the gaseous disks are dissipated, which have dramatically changed the system architectures \citep{Rasio:1996, Weidenschilling:1996, Lin:1997}. Population syntheses have indeed been able to reproduce the observed eccentricity distribution of giant exoplanets, starting from quite different initial configurations (e.g., \citealt{Chatterjee:2008, Juric:2008, Ford:2008, Ida2013, Carrera:2019, Bitsch:2020, Bitsch:2023, Nagpal:2024, Pan:2025}). A comprehensive review of the relevant literature studies is beyond the scope of the present work. Instead, we choose to focus our discussion on a few representative studies.

\citet{Juric:2008} performed N-body simulations with initially different numbers (3, 10, and 50) of planets at random locations. While they were able to reproduce well the eccentricity distribution of extra-solar giant planets, they predicted that each system should have at least two survival planets with masses above $\sim$0.3\,$M_{\rm J}$
\footnote{Although their simulation setup contains planets down to $0.1\,M_{\rm J}$, the lower-mass ones are largely removed/absorbed during the dynamical evolution.}
and semi-major axis within $\sim$100\,au. 
The spatial range of their planets is wider, but the number within $\sim30\,$au is not expected to change too much: The two (if not more) remaining planets are commonly separated by $\sim$15 mutual Hill radius, so if one planet is at $\lesssim5\,$au, the other would be $\lesssim13\,$au, which is still within the window of semi-major axis that is under study. Therefore, the simplistic initial conditions of \citet{Juric:2008} predicted too many giant planets.

Another influential study is \citet{Chatterjee:2008}, which started with three giant planets in initially packed configurations. At the end of their orbital integrations, all have lost at least one planet, with 20\% having lost two, resulting an average multiplicity of the final giant planet systems of $1.8$. This seems to agree well with our derived value (after including the linear trends). However, \citet{Chatterjee:2008} predicted that the the number of two-giant-planet systems exceeds the number of one-giant-planet systems by a factor of four, which is much higher than our derived ratio ($F_2/F_1\approx 1.0$ for CLS and $0.3$ for HARPS).

More recent studies involving the giant planet--giant planet scatterings have included formation and migration phases of the planets, the addition and mutual impact of the lower-mass companions, and/or the tailored prescriptions to reproduce specific giant planet populations/properties \citep[e.g.,][]{Wu:2011, Frelikh:2019, Carrera:2019, Bitsch:2023, Nagpal:2024, Pan:2025}. One common outcome of these studies is that they tend also to produce too many multi-giant systems similar to the early works. It is interesting to note that certain simulations in \citet{Nagpal:2024} do seem to reproduce the multiplicity distribution well, but the required parameters point to very weak eccentricity damping from the gaseous disk (see also \citealt{Bitsch:2020}). Both \citet{Nagpal:2024} and \citet{Bitsch:2020} adopted the \citet{Lee:2002} prescription of eccentricity damping, but the full details of this damping is much more complicated. More theoretical investigations are needed to better understand the dynamical evolution of giant planets in the gaseous environment and how it affects the evolution of planet multiplicity.

\subsection{Correlation between Different Planet Populations}

The occurrence rate of all giants is noticeably lower than the direct sum of the close-in and cold-giant occurrence rates, suggesting that these two types of giants may seem to co-exist. To test this, we follow \citetalias{Zhu:2022} and consider the null hypothesis in which the presence of close-in and cold giants is uncorrelated. Under this assumption, the expected fraction of systems hosting at least one giant planet within 10\,au is $12.0\%$ ($20.5\%$) for HARPS (the refined CLS), which is higher than the directly derived value of $10.6\%$ ($15.8\%$). This suggests that the two populations are probably correlated.

To go one step further, we can derive the conditional rate, $P({\rm cold}|{\rm close})$, which is the frequency of systems with cold Jupiters given that there are already inner hot/warm Jupiters. This conditional rate is given by
\begin{equation}
    P({\rm cold}|{\rm close}) = \frac{P({\rm close}) + P({\rm cold}) - P({\rm all})} {P({\rm close})} .
\end{equation}
This conditional rate is $\sim50\%$ for HARPS and $\sim100\%$ for the refined CLS. Although these derived rates are not statistically significant by themselves, this method can potentially provide an independent check for the inner--outer correlation of giant planets \citep{Knutson:2014, Bryan:2016}.

\begin{acknowledgments}
%We would like to thank Fei Dai and Scott Tremaine for comments and suggestions on an earlier version of the manuscript. 
This work is supported by the National Science Foundation of China (grant No.\ 12173021 and 12133005). We thank Eugene Chiang for early discussions and the anonymous reviewer for comments on the manuscript.
\end{acknowledgments}

\bibliography{my_bib}{}
\bibliographystyle{aasjournalv7}

%% This command is needed to show the entire author+affiliation list when
%% the collaboration and author truncation commands are used.  It has to
%% go at the end of the manuscript.
%\allauthors

%% Include this line if you are using the \added, \replaced, \deleted
%% commands to see a summary list of all changes at the end of the article.
%\listofchanges

\end{CJK*}
\end{document}